\documentclass[manuscript]{aastex}

\usepackage{floatrow}

\shorttitle{GCE AND SOLAR S-PROCESS ABUNDANCES}
\shortauthors{Bisterzo et al.}

\begin{document}

\title{GALACTIC CHEMICAL EVOLUTION: THE IMPACT OF THE $^{13}$C-POCKET
STRUCTURE ON THE S-PROCESS DISTRIBUTION}

\author{S. Bisterzo\altaffilmark{1} and C. Travaglio\altaffilmark{2}}
\affil{INAF - Astrophysical Observatory Turin, Turin, Italy}

\email{bisterzo@to.infn.it; sarabisterzo@gmail.com}

\author{M. Wiescher} 
\affil{Joint Institute for Nuclear Astrophysics (JINA), Department 
of Physics, University of Notre Dame, IN, USA}

\author{F. K{\"a}ppeler}
\affil{Karlsruhe Institute of Technology, Campus Nord, 
Institut f{\"u}r Kernphysik, Karlsruhe, Germany}

\and

\author{R. Gallino\altaffilmark{2}}
\affil{Department of Physics, University of Turin, Italy}

\altaffiltext{1}{Department of Physics, University of Turin, Italy}

\altaffiltext{2}{B2FH Association-c/o Strada Osservatorio 20, 10023 Turin, Italy}

\begin{abstract}
The solar $s$-process abundances have been analyzed in the framework of a Galactic Chemical
Evolution (GCE) model.
The aim of this work is to implement the study by \citet{bisterzo14}, who investigated
the effect of one of the major uncertainties of asymptotic giant branch (AGB) yields, 
the internal structure of the $^{13}$C pocket.
We present GCE predictions of $s$-process
elements computed with additional tests 
in the light of the suggestions provided in recent publications. 
\\
The analysis is extended to different metallicities, by comparing GCE results and
updated spectroscopic observations of unevolved field stars. 
We verify that the GCE predictions obtained with different tests may represent, 
on average, the evolution of selected neutron-capture elements in the Galaxy.
The impact of an additional weak $s$-process contribution from fast-rotating 
massive stars is also explored.

\end{abstract}



\section{GCE Solar s-process Predictions}\label{intro}

AGB stars with low initial mass are the major responsible for the nucleosynthesis
of solar $s$ isotopes with $A$ $>$ 90 \citep{busso99}.
\\
The main neutron source of low-mass AGB models is the $^{13}$C($\alpha$, n)$^{16}$O reaction,
which burns radiatively during the interpulse in a thin layer of the He intershell, the 
so-called {\it $^{\it 13}$C} $pocket$ \citep{straniero95}. 
The formation of the $^{13}$C pocket requires an unknown mixing mechanism that allows
partial mixing of a few protons from the convective envelope into the top 
layers of the radiative He- and C-rich intershell. This is assumed to occur
at the quenching of a Third Dredge Up (TDU) episode.
When the star contracts, the H shell reignites and protons in the intershell 
are captured by the abundant $^{12}$C nuclei to yield primary $^{13}$C.
If more protons than $^{12}$C nuclei are diffused in the outer layers, a region of primary
$^{14}$N may form by further proton captures on $^{13}$C.
Subsequently, the temperature in the $^{12}$C pocket increases to
$\sim$1$\times$10$^8$ K, and neutrons are released radiatively within the pocket via 
$^{13}$C($\alpha$, n) reactions at 
quite low neutron densities.

Various physical mechanisms have been explored for the formation of $^{13}$C pocket
(e.g., overshooting, rotation, magnetic fields, gravity waves; 
\citealt{herwig97,langer99,denissenkov03,siess04,straniero06,piersanti13,nucci14}).
The details of how the $^{13}$C pocket forms are still debated, making its mass extent
and the H profile largely uncertain.
\\
Spectroscopic observations provide key information to constrain theoretical models:
chemically peculiar $s$-rich stars have evidenced a dispersion of the $s$ abundances
for a given metallicity (e.g., MS, S, C(N), Ba, CEMP-s and post-AGB stars, planetary 
nebulae; see the recent review by \citealt{kaeppeler11} and \citealt{karakas14}). 
This dispersion 
has been recognized since first studies by \citet{busso01} and \citet{abia02}, but
the reason(s) are not definitely identified.
A variation in the stellar rotational velocity may be regarded as a possible explanation 
(see \citealt{piersanti13}, and references therein).

Owing to the present uncertainties, the $^{13}$C pocket is artificially introduced
in our post-process AGB models, following the observational constraints.
The shape of the $^{13}$C 
and $^{14}$N profiles and the mass involved in the pocket are regulated by 
a free parametrization. 
The internal structure of the $^{13}$C pocket adopted so far has been calibrated to 
represent the solar main component \citep{arlandini99}: it is a three-zone pocket
(each zone has defined $X$($^{13}$C) and $X$($^{14}$N) abundances) with a total mass 
of about 0.001 $M_\odot$ (see Table~\ref{Tab1}; first group of data).  
A range of $^{13}$C-pocket strengths is assumed to reproduce the spectroscopic 
$s$-process dispersion: 
we parametrically vary the concentration of $^{13}$C (and $^{14}$N) of each zone 
given in Table~\ref{Tab1} by different factors,
leaving the mass of the pocket constant. 
We refer to \citet{bisterzo10,bisterzo14} for a detailed discussion.
\\
This systematic approach appears justified by the present 
uncertainties: the formation of diverse $^{13}$C pockets may result from 
the interplay between different physical processes in stellar interiors.
Post-process models should be considered 
 useful tests to address full stellar evolutionary models (and in general 
multidimensional/hydrodynamical simulations) against observational constraints.

The solar $s$-process abundances must account for the complex chemical evolution 
of the Galaxy, which includes AGB yields of various masses and metallicities.
The chemical evolution model adopted to reproduce the solar $s$ distribution has been
exhaustively described by \citet{travaglio99,travaglio04}. 
\\
In the framework of GCE, we showed that the solar $s$ distribution of isotopes with 
130 $<$ $A$ $\leq$ 208 can be accurately reproduced once we consider 
a proper weighted average among the $^{13}$C-pocket strengths\footnote{To this purpose, the 
unbranched $s$-only 
isotope $^{150}$Sm is taken as reference nuclide for the whole $s$-process distribution.
The high production of $^{208}$Pb in low-mass low-metallicity AGB stars plays
another indicative GCE constraint \citep{travaglio01}.
The solar abundance of $^{208}$Pb is matched once the $s$-process occurring in low-metallicity 
AGB stars is properly considered in the context of the chemical evolution of the Galaxy.}. 
This is consistent with the observed spectroscopic $s$-process dispersion.
A deficit (of about 25\%) between GCE predictions of $s$-process
elements and the solar abundances was found for
isotopes with 90 $<$ $A$ $<$ 130 (solar LEPP-s).
\citet{bisterzo14} have investigated a possible connection between this deficit and the 
$^{13}$C-pocket structure. On the basis of their tests, solar GCE predictions of $s$-process
elements are marginally affected. 

The aim of this work is to implement the analysis carried out on the 
$^{13}$C-pocket structure by \citet{bisterzo14}, (Section~\ref{tests}).
In Section~\ref{yields}, we consider the sensitivity of AGB yields 
to metallicity, focusing on the
contribution by metal-rich AGB stars to the 
light elements (see discussion by \citealt{maiorca12} for open
clusters). 
Although the solar composition is fundamental to constrain AGB yields, it
only provides a single piece of information about the Galactic history. 
The reliability of the $^{13}$C-pocket tests needs to be verified by 
considering the complex framework of Galactic chemical enrichment.
The GCE predictions of selected neutron-capture elements versus metallicity 
are compared with updated spectroscopic observations in Section~\ref{z}.
Recently, rotation-induced mixing in low-metallicity massive stars
has been proposed as an explanation of the observed [Sr/Ba] 
dispersion in extremely metal-poor stars, being efficient primary producers for 
$s$ isotopes heavier than Sr, up to the Ba neutron-magic peak
\citep{fris12,pignatari13,cescutti13}.
In this context, we investigate the impact of recent weak $s$-process
yields by \citet{fris16}, available from a large grid of rotating massive stars 
($Z$ from 10$^{-5}$ to solar), on the Galactic chemical enrichment (Section~\ref{rot}).
\\
In Section~\ref{conclusions} our results are briefly summarized and future outlook 
are discussed.


\section{Impact of New $^{13}$C-pocket Tests on Solar $s$ Abundances}\label{tests}

We focus on specific additional tests carried out 
on the basis of recent advice available in literature.
The internal $^{13}$C-pocket structure adopted in each test is given in 
Table~\ref{Tab1}\footnote{We remind that all tests are performed on low-mass AGB models
because the effect of the $^{13}$C pocket in AGB stars with intermediate mass 
(4 $\leq$ $M$/$M_\odot$ $<$ 8) is negligible for GCE solar predictions (see 
Section~2 by \citealt{bisterzo14}; \citealt{straniero14}).}. 
\begin{itemize}
\item Starting from the three-zone $^{13}$C-profile adopted so far, we investigate the impact 
of a substantially extended $^{13}$C-pocket mass than assumed in our previous computations
(up to four times larger, corresponding to a total mass of $M_{\rm tot}$(pocket)$\sim$4$\times$10$^{-3}$ $M_\odot$;
see tests described as {\bf CASE 1} and {\bf CASE 2} in Table~\ref{Tab1}).
\\
In this regard, \citet{maiorca12} have proposed such a $^{13}$C-pocket mass to represent the 
abundances of neutron-capture elements in young open clusters. Magnetic buoyancy (or other forced
mechanisms) are suggested as good candidates to form such a $^{13}$C reservoir \citep{trippella14}. 
Rotation models by \citet{piersanti13} indicate that 
low-metallicity AGB stars 
and fast-rotating metal-rich stars might produce such an extended $^{13}$C pocket. 
Comparison between theoretical models and the strontium and barium isotopic signatures 
measured in mainstream SiC grains require $^{13}$C pockets with 
$M_{\rm tot}$(pocket)$\geq$1$\times$10$^{-3}$ $M_\odot$
\citep{liu15}.
\item In {\bf CASE 3} (see Table~\ref{Tab1}), we test the effect of an additional parametrized 
$^{14}$N-pocket.
\\
As anticipated in Section~\ref{intro}, an outer $^{14}$N-rich layer may form in the 
pocket once enough protons are mixed in the external zone of the He intershell
\citep{goriely00,cristallo09,cristallo11,karakas10,lugaro12,trippella14}.
The presence of rotation may widen the 
$^{14}$N pocket by inducing partial mixing of the $^{14}$N-rich region with the inner 
$^{13}$C-rich zone.
\item In {\bf CASE 4}, we have performed a set of 1.5, 2 and 3 $M_\odot$ models 
computed with a more efficient Reimers mass loss than in previous models.
This implies that AGB models experience about half of the TDU episodes compared to CASE 3.
The $^{13}$C pocket structure is the same as adopted in CASE 3 
(see Table~\ref{Tab1}). 
\\
This test simulates the recent prescriptions of
updated FRUITY\footnote{web: fruity.oa-teramo.inaf.it/.} models by \citet{cristallo09,cristallo11}.
FRUITY models include an efficient AGB mass-loss rate, which has been calibrated 
using the infrared luminosity function of Galactic carbon stars, 
and improved radiative C-enhanced opacity tables. 
Accordingly, these new models with a reduced number of TDU episodes are  
in better agreement with observations 
(\citealt{guandalini13} and references therein). 
\item Finally, in {\bf CASE 5} (see Table~\ref{Tab1}), we test the impact of an additional extended
inner zone of the pocket with a mass of 2$\times$10$^{-3}$ $M_\odot$ and a correspondingly lower
$^{13}$C abundance ($X$($^{13}$C)=2.75E$-3$).
\\
Recently,
\citet{cristallo15} 
found that a different convective/radiative boundary condition 
allows a deeper penetration of protons with a very low mixing efficiency during TDU episodes.
The resulting $^{13}$C pocket displays an extended tail with a smooth decrease of the $^{13}$C 
profile (see $Tail$ model in their Fig.~7). 
CASE 5 roughly approximates the $Tail$ model by \citet{cristallo15}.
\end{itemize}

\begin{deluxetable}{ccccccccc}
\tabletypesize{\scriptsize}
\tablewidth{0pt}
\tablecaption{\label{Tab1} Internal structure of the $^{13}$C-pockets adopted in
the tests displayed in Fig.~\ref{Fig1}.} 
\tablecolumns{9}          
\tablehead{    &    &  Zone 0      &  Zone I    &  Zone II   & Zone III  &  Zone IV & Zone V  } 
\startdata  
\multicolumn{9}{l}{{\bf Standard choice}: three-zone model; $M_{\rm tot}$(pocket) = 1.09E$-$3 $M_\odot$} \\
&Mass($M_\odot$)  & $-$  & 5.50E$-$4    &  5.30E$-$4   &   7.50E$-$6 & $-$  & $-$  &  \\   
&$X$($^{13}$C)  &  $-$ & 3.20E$-$3    &  6.80E$-$3   &   1.60E$-$2 & $-$  &  $-$   &  \\ 
&$X$($^{14}$N)  &  $-$ & 1.07E$-$4    &  2.08E$-$4   &   2.08E$-$3 & $-$  &  $-$   &  \\ 
\cline{1-9}
\multicolumn{9}{l}{{\bf CASE 1}$^a$: three-zone model; $M_{\rm tot}$(pocket) = 3$\times$1.09E$-$3 $M_\odot$} \\ 
&Mass($M_\odot$)  & $-$  & 1.65E$-$3    &  1.59E$-$3   &   2.25E$-$5 & $-$  & $-$  &   \\   
&$X$($^{13}$C)  &  $-$ & 3.20E$-$3    &  6.80E$-$3   &   1.60E$-$2 & $-$  &  $-$   &  \\  
&$X$($^{14}$N)  &  $-$ & 1.07E$-$4    &  2.08E$-$4   &   2.08E$-$3 & $-$  &  $-$   &  \\    
\cline{1-9}
\multicolumn{9}{l}{{\bf CASE 2}$^b$: 1.3 $\leq$ $M$ $\leq$ 1.5 $M_\odot$ models; $M_{\rm tot}$(pocket) = 4$\times$1.09E$-$3 $M_\odot$}  \\ 
&Mass($M_\odot$)  & $-$  & 2.20E$-$3    &  2.12E$-$3   &   3.00E$-$5 & $-$  &  $-$  &   \\   
&$X$($^{13}$C)  &  $-$ & 3.20E$-$3    &  6.80E$-$3   &   1.60E$-$2 & $-$  &  $-$    &  \\  
&$X$($^{14}$N)  &  $-$ & 1.07E$-$4    &  2.08E$-$4   &   2.08E$-$3 & $-$  &  $-$    &  \\         
\cline{1-9}
\multicolumn{9}{l}{{\bf CASE 3 and CASE 4}$^c$: five-zone model with external $^{14}$N-rich zones; $M_{\rm tot}$(pocket) = 3.24E$-$3 $M_\odot$}\\ 
&Mass($M_\odot$)  & $-$  & 1.20E$-$3    & 1.05E$-$3    & 3.30E$-$4  & 3.30E$-$4 & 3.30E$-$4 & \\ 
&$X$($^{13}$C)  &  $-$ & 3.20E$-$3    &  6.80E$-$3   &   1.60E$-$2 & 4.00E$-$2 & 4.00E$-$2  &\\ 
&$X$($^{14}$N)  &  $-$ & 1.07E$-$4    &  2.08E$-$4   &   2.08E$-$3 & 8.00E$-$2 & 1.49E$-$1 & \\ 
\cline{1-9}
\multicolumn{9}{l}{{\bf CASE 5}$^d$: four-zone model with an extended inner tail of $^{13}$C; $M_{\rm tot}$(pocket) = 3.09E$-$3 $M_\odot$}  \\ 
&Mass($M_\odot$)  & 2.00E$-$3  & 5.50E$-$4    &  5.30E$-$4   &   7.50E$-$6   & $-$  & $-$&    \\   
&$X$($^{13}$C)  & 2.75E$-$3  & 3.20E$-$3    &  6.80E$-$3   &   1.60E$-$2   &  $-$ &  $-$ &    \\ 
&$X$($^{14}$N)  & 5.73E$-$5  & 1.07E$-$4    &  2.08E$-$4   &   2.08E$-$3  &  $-$ &  $-$  &  \\ 
\cline{1-9}
\multicolumn{9}{l}{$^a$ The mass of the pocket is increased by a factor of three with respect to our standard choice
 \citep{bisterzo14}.}\\
\multicolumn{9}{l}{$^b$ The mass of the pocket is increased by a factor of four with respect to our standard choice in}\\
\multicolumn{9}{l}{AGB models of 1.3 to 1.5 $M_\odot$. We leave the $^{13}$C pocket unchanged for $M$ $>$ 1.5 $M_\odot$ models.}\\
\multicolumn{9}{l}{$^c$ We include two additional external zones with $X$($^{14}$N) much higher than $X$($^{13}$C).}\\
\multicolumn{9}{l}{$^d$ We include an additional internal zone with $X$($^{13}$C) = 2.75E$-3$.}\\
\enddata                         
\end{deluxetable}

\begin{figure} 
\center
\includegraphics[angle=-90,width=16cm]{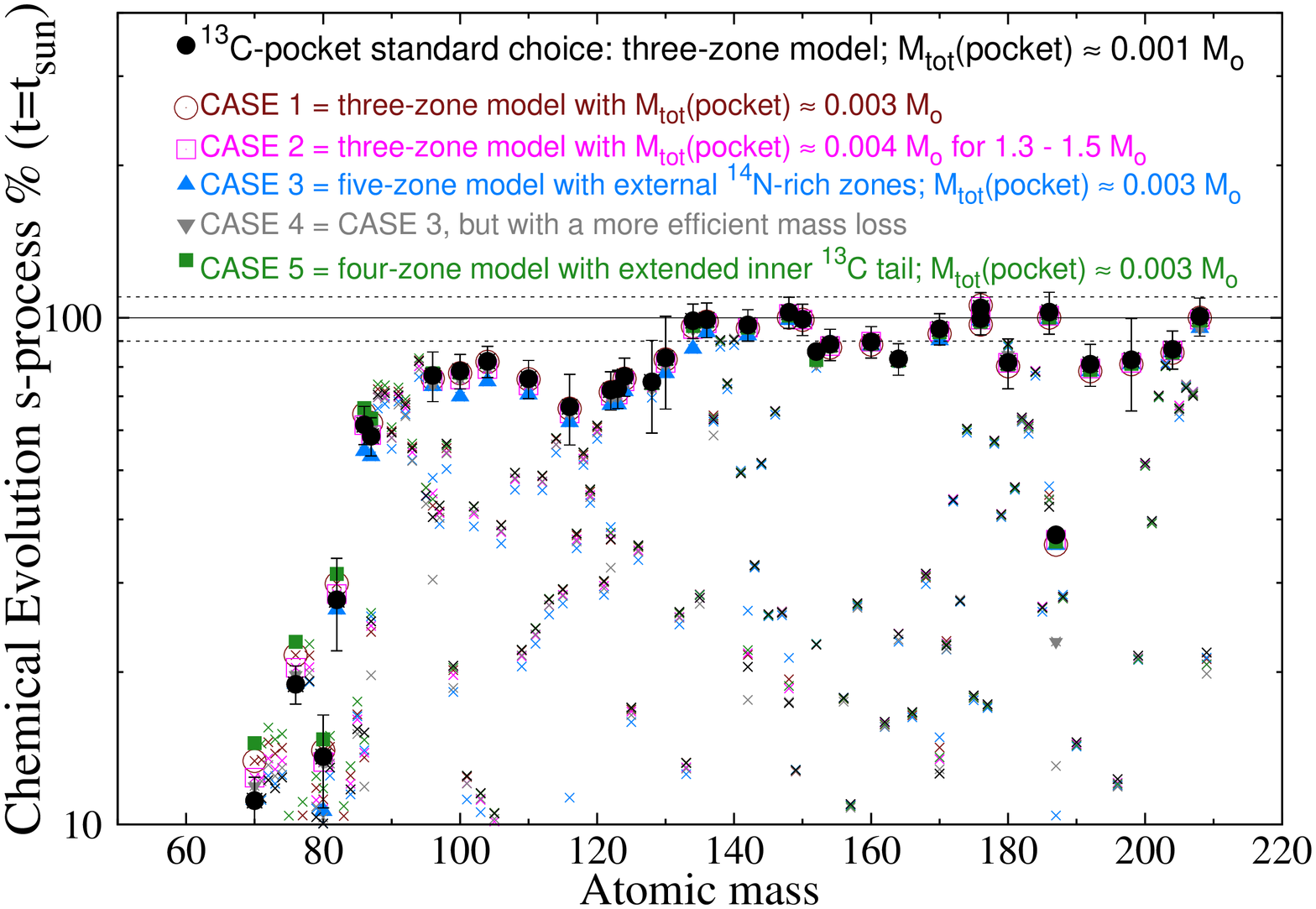}
\caption{\label{Fig1}
Effect of the $^{13}$C-pocket uncertainties in low mass AGB models on GCE solar $s$-process 
predictions. The $s$-only isotopes (and $^{208}$Pb) obtained with our standard three-zone 
$^{13}$C-pocket choice are represented by filled circles. 
Stable nuclei are displayed by crosses.
The results obtained by several tests have been displayed with different symbols
(see label in the top panel of the figure). 
We have adopted a range of $^{13}$C-pocket strengths (Table~\ref{Tab1})
in order to reproduce 100\% of solar $^{150}$Sm when 
changing the structure of the $^{13}$C pocket (see \citealt{bisterzo14}, their Fig.~4).
Note that $s$-only isotopes with $A$ $<$ 90 ($^{70}$Ge, $^{76}$Se, $^{80,82}$Kr, 
$^{86,87}$Sr) receive an additional contribution by the weak $s$-process in
massive stars (see e.g., \citealt{pignatari10}). }
\end{figure}

The resulting solar GCE predictions of $s$-process
elements are displayed in Fig.~\ref{Fig1}.
A proper weighted average among the various $^{13}$C-pocket strengths must 
be adopted for each test to reconcile GCE predictions with 100\% of solar $^{150}$Sm
(see \citealt{bisterzo14}).  
This approach allows us to reproduce the solar $^{150}$Sm within 5\% uncertainty
\citep{lodders09}. 
\\
The $s$-only isotopes with atomic mass $A$ $\ga$ 90 show variations 
smaller than $\sim$10\%, 
thus confirming the need of a solar LEPP-s mechanism in order to increase 
the solar $s$ abundances in the range 90 $<$ $A$ $<$ 130, as predicted by \citet{travaglio04}.

Note that AGB yields computed with a single $^{13}$C-pocket choice do not provide accurate 
 interpretations 
neither of solar $s$ abundances nor of observations of peculiar $s$-rich stars.
By working in a larger range of uncertainties, potential missing contributions are not 
necessarily highlighted.
Accordingly, the result of this paper does not disagree with a recent study by \citet{cristallo15} 
that provides a meticulous discussion about the uncertainties affecting stellar models and
the solar GCE distribution.
Within the estimated uncertainties a LEPP mechanism is not necessarily required.
However, the authors suggest how their representation of the solar distribution could be 
improved once models with different initial rotational velocities (or different 
prescriptions for convective overshoot during the TDU) 
will be included in GCE computations
for an extended metallicity grid.
These upcoming progresses in stellar models will assess whether additional 
contributions are needed.

\section{The Stellar Yields versus Metallicity}\label{yields}

\begin{figure} 
\vspace{-0.7cm}
\includegraphics[angle=0,width=9.7cm]{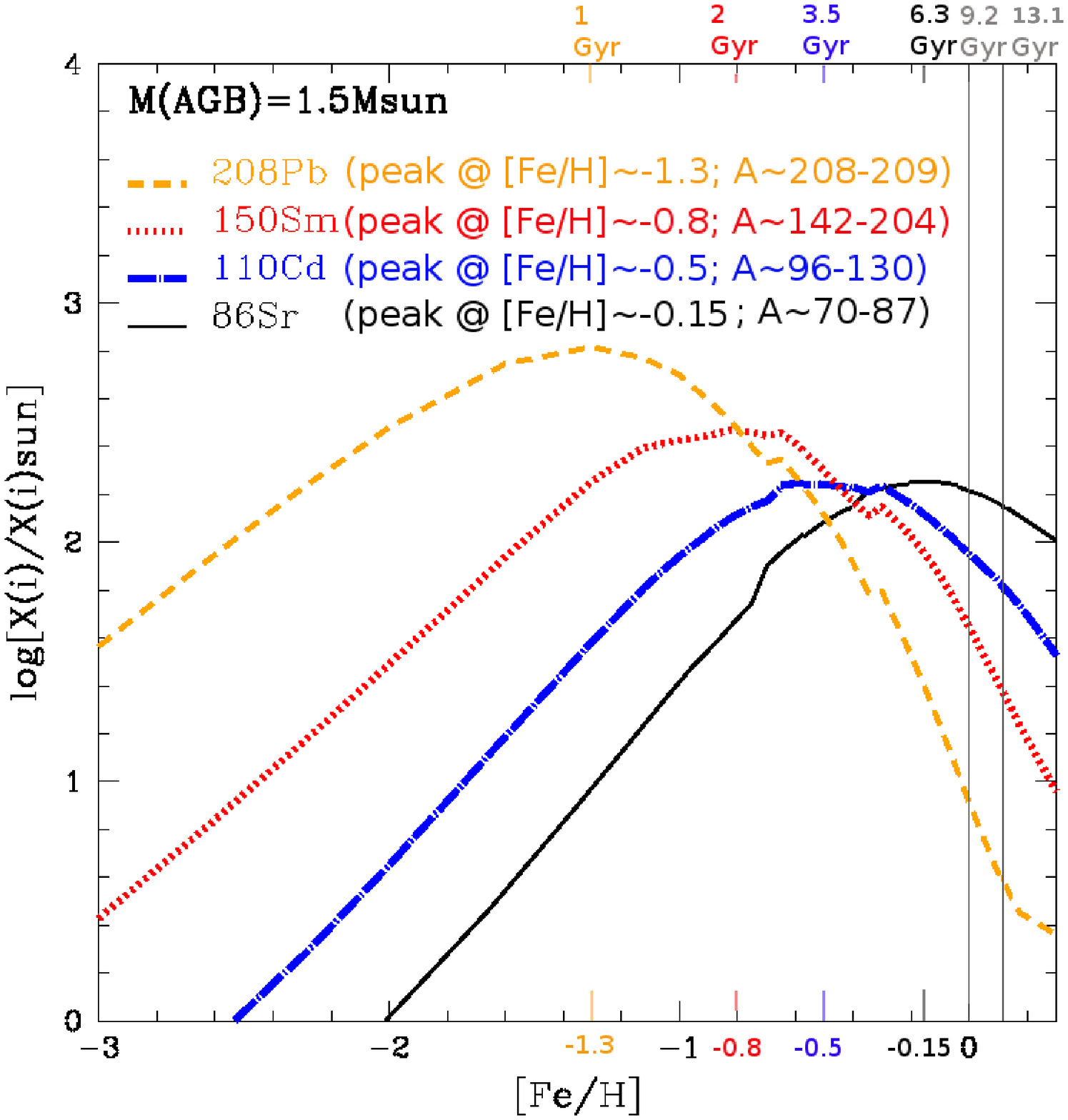} 
\vspace{-0.5cm}
\includegraphics[angle=-90,width=9cm]{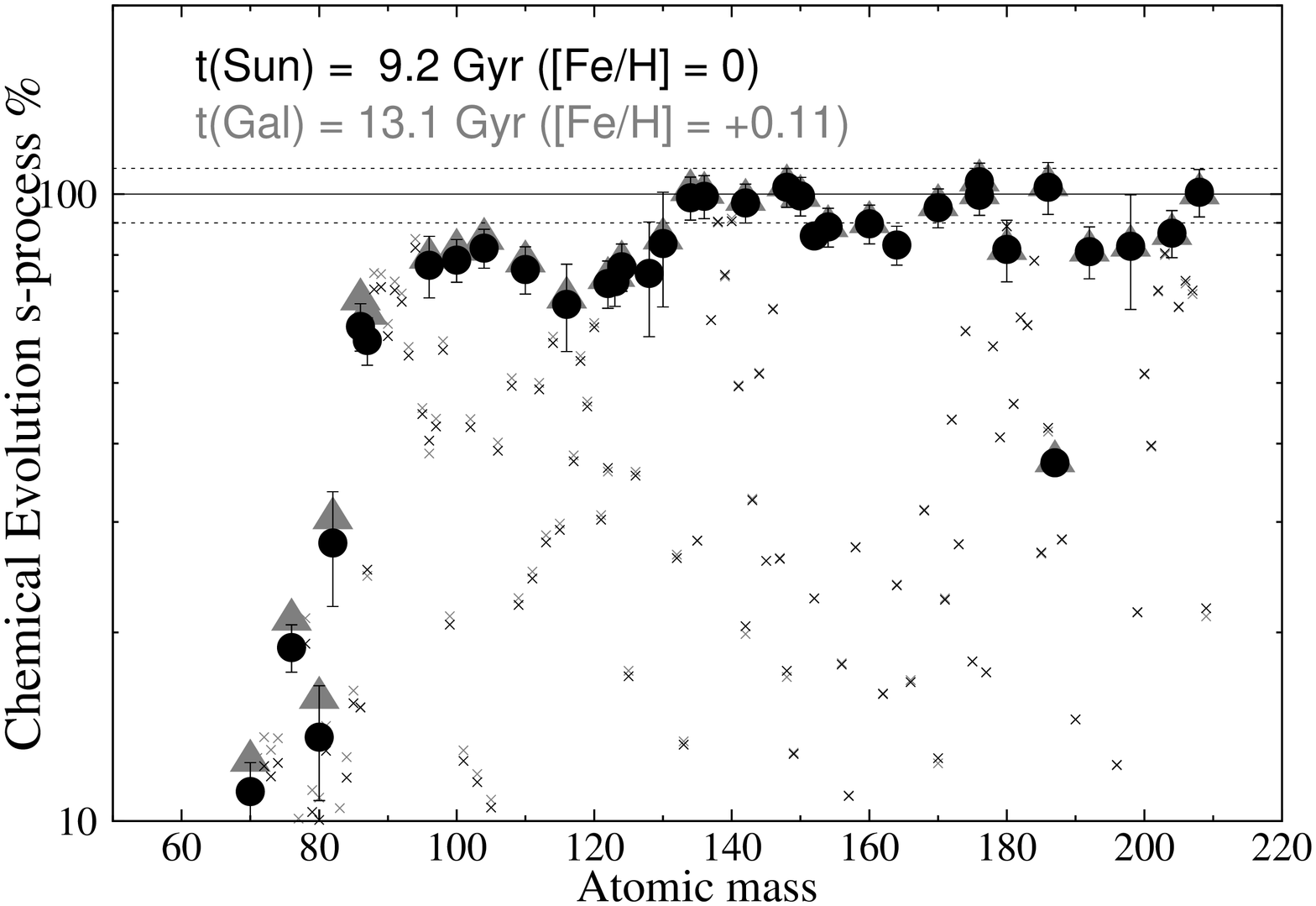} 
\caption{\label{Fig2}
Top panel: AGB yields (obtained with a weighted average of standard $^{13}$C-pockets) of a 1.5 $M_\odot$ 
model at different metallicities. Four selected $s$-only isotopes are shown: 
$^{150}$Sm is representative of heavy $s$-only isotopes with $A$ = 142 $-$ 204 (peaked at
[Fe/H] $\sim$ $-$0.8, which in our GCE model corresponds to a Galactic time of $t_{Gal}$ $\sim$ 2 Gyr); 
$^{110}$Cd illustrates the behavior of light $s$-only isotopes with
$A$ = 96 $-$ 130 ([Fe/H] $\sim$ $-$0.5; $t_{Gal}$ $\sim$ 3.5 Gyr); last are the $s$-only isotopes with 
$A$ = 70 $-$ 87, represented by $^{86}$Sr ([Fe/H] $\sim$ $-$0.15; $t_{Gal}$ $\sim$ 6.3 Gyr).
 $^{208}$Pb is efficiently produced by low-metallicity AGB 
stars ([Fe/H] $\sim$ $-$1.3; $t_{Gal}$ $\sim$ 1 Gyr).
Bottom panel: the $s$-process distribution obtained at the epoch of the solar 
system formation ($t_\odot$ = 9.2 Gyr at [Fe/H] = 0; black symbols) compared with the $s$-process distribution 
obtained at a Galactic time of $t_{Gal}$ = 13.1 Gyr (at [Fe/H] = +0.11; gray symbols). Both results have been
normalized to $^{150}$Sm. }
\end{figure}

The complex dependence of $s$-process yields on the initial stellar metallicity 
is well known.
In Fig.~\ref{Fig2} (top panel), we display the AGB production factors of selected 
$s$-only isotopes. 
Starting from [Fe/H] = $-$0.5, which corresponds to $t_{Gal}$ = 3.5 Gyr, the 
s contribution to light isotopes with A = 96 -- 130 is favored. 
As suggested by \citet{maiorca12} and \citet{cristallo15},
it is crucial to verify that the delayed contribution of metal-rich AGB stars to the solar 
epoch is properly considered in the framework of our GCE model.
Otherwise, the solar LEPP-s may derive from a relevant underestimation of the 
metal-rich AGB component.
\\
 The theoretical minimum AGB initial mass that can contribute 
to the solar system is rather uncertain and model dependent. In our GCE computations,
$M$ $\le$ 1.3 $M_\odot$ models with [Fe/H] $\ga$ $-$0.4 
do not contribute to the chemical evolution of the Galaxy because 
the conditions for the activation of the TDU episodes are never reached in our models
(see e.g., \citealt{bisterzo14}, and references therein). 
\\
We have analyzed here a test case, in which
we have accounted for the longest life of the AGB stars that may contribute to the solar
distribution (e.g., our 1.3 $M_\odot$ models with disc metallicity 
have lifetime of $\sim$4 Gyr).
In Fig.~\ref{Fig2} (bottom panel), we compare the $s$ distribution obtained at the formation 
time of the solar system ($t_\odot$ = 9.2 Gyr) and that computed at $t_{Gal}$ = 13.1 Gyr,
in which the overall metal-rich AGB component that may contribute to the solar s distribution
is included.
The two distributions show marginal variations ($\la$3\%) for $A$ $>$ 90. 
Indeed, our GCE model adopts a very efficient star formation rate in the halo, which 
produces a fast metallicity increase at the early evolutionary epochs ([Fe/H] = $-$2 at 0.1 
Gyr). An extended thin disc phase ($\sim$8 Gyr) assures that the contribution of metal-rich stars 
is not overlooked. 
Moreover, the small star formation rate after 9.2 Gyr results in a slow metallicity 
increase (by +0.11 dex from solar to 13.1 Gyr). For this reason, the solar distribution 
(normalized to $^{150}$Sm) shows small variations at the two selected epochs.
This indicates that the solar distribution properly accounts for the contribution of AGB 
yields with [Fe/H] $\la$ $-$0.2.
\\
Non-negligible variations are found for $^{86,87}$Sr and
$^{89}$Y: at $t_{Gal}$ $\sim$13 Gyr, $\Delta$($^{86,87}$Sr)$\simeq$+8\% and $\Delta$($^{89}$Y)$\simeq$+4\%.
The $s$-only isotopes with $A$ $<$ 90 are mainly produced by AGB stars with [Fe/H]$\sim$$-$0.15, which 
corresponds to $t_{Gal}$ $\sim$ 6.3 Gyr, 
only about 2.9 Gyr before the solar system formation (see Fig.~\ref{Fig2}, top panel). 
This is comparable to the 
lifetime of AGB stars with $M$ = 1.4 $-$ 1.5 $M_\odot$ with [Fe/H]$\sim$$-$0.15
($t_{\star}$ $\sim$ 3.0 to 2.4 Gyr, respectively).
\\
In Table~\ref{Tab2}, we list the $s$ contributions to elements from Sr to Cd at $t_\odot$ = 9.2 Gyr
and $t_{Gal}$ $\sim$13 Gyr: although non-negligible variations are found for Sr (+5\%) and Y (+4\%), 
these differences are compatible with the solar uncertainties. As
discussed in the previous Section, the LEPP-s is still required.

\begin{table}
\caption{The $s$ contributions to elements from
Sr to Cd at $t_\odot$ = 9.2 Gyr and $t_{Gal}$ = 13.1 Gyr.}\label{Tab2}
\centering
\resizebox{8cm}{!}{\begin{tabular}{cccc}
\hline
Element   & $t_\odot$ = 9.2 Gyr    &  $t_{Gal}$ = 13.1 Gyr  &  $t_{Gal}$ $-$ $t_\odot$ \\
\hline
Sr &    69  &  74      &  5   \\   
Y  &    72  &  76      &  4   \\   
Zr &    66  &  68      &  2   \\   
Nb &    56  &  58      &  2   \\   
Mo &    39  &  40      &  1   \\   
Ru &    29  &  30      &  1   \\   
Rh &    12  &  12      &  0   \\   
Pd &    36  &  37      &  1   \\   
Ag &    11  &  11      &  0   \\   
Cd &    46  &  47      &  1   \\   
\hline    
\end{tabular}}
\end{table}

\section{GCE $s$- and $r$-process contributions versus {[Fe/H]} }\label{z}

In this section we present our results for the chemical evolution of selected
neutron-capture elements as a function of metallicity.
We analyze the chemical evolution of Y, Zr, Ba, La and Pb (representative 
of the three $s$-process peaks), and Eu (typical of the $r$-process elements).
GCE predictions are compared with updated high-resolution
spectroscopic observations.

The Galactic enrichment of the three $s$-process peaks is followed by 
accounting for the $s$-process AGB yields discussed in Sections~\ref{tests}
and~\ref{yields}.
An additional contribution to the first $s$ peak is ascribed to the 
`classical' weak $s$ component from (non-rotating) massive stars 
\citep{raiteri92,pignatari10}. Given the secondary-like nature of the 
$^{22}$Ne($\alpha$, n)$^{25}$Mg neutron source in standard non-rotating 
massive stars (with yields scaling quite linearly with the initial metallicity), 
the weak $s$ process is mainly relevant at solar metallicity, accounting
for up to $\sim$10\% of solar Sr-Y-Zr \citep{travaglio04}. 
We will discuss the supplementary impact of fast-rotating low-metallicity massive
stars on GCE predictions in Section~\ref{rot}. 
\\
Concerning the treatment of Eu and $r$-process elements, we have adopted
the same prescription discussed by \citet{travaglio99}. 
Actually, the origin of the $r$ process is still unclear and largely uncertain.
Spectroscopic observations have evidenced the existence of multiple $r$-process 
components operating in the early Galaxy (see, e.g., 
\citealt{siq15,roederer14,hansen14,hansen14a,hansen12,peterson13,montes07}, and references therein). 
At least two $r$-processes have been {bf\ proposed}:
the main $r$ process ($A$ $>$ 130), which is currently ascribed to mergers of 
compact objects and/or magneto-rotationally driven supernovae, 
and the weak $r$ process (or $\nu$p-process, rp-process; $A$ $\sim$ 80 to 120), 
which has been proposed as an alternative nucleosynthesis source of lighter heavy elements 
in neutrino-driven supernova outflows.
Noteworthy progresses have been reached in modeling these extremely energetic 
and complex events,
but the unclear physical conditions of the astrophysical sites 
and the inaccessibility of nuclear physics inputs lead to very uncertain
$r$-process predictions (see e.g., 
\citealt{pereiramontes16,eichler15,goriely15,nishimura15,kratz14,winteler12,arcones13,qianwass08}, 
and references therein).
\\
The lack of a comprehensive $r$-process view forced us to systematically include 
an $r$-process contribution in GCE computations. 
For elements heavier than Ba, the solar $r$-process contribution is derived
by subtracting the $s$ fractions from the solar abundances (the so-called $r$-process 
residuals method; \citealt{kaeppeler82,kaeppeler11}). 
Following the observed decreasing trend of heavy neutron-capture elements in the early 
Galaxy, we ascribe the $r$ contribution to a primary process occurring in SNII with a 
limited range of progenitor masses, $M$ $\sim$ 8$-$10 $M_\odot$ \citep{travaglio99}.
In light of the present uncertainties,
we do not exclude different hypotheses (e.g., Jet-SNe, neutron star mergers).
Chemo-dynamical evolution models are needed to adequately explore the chemical 
origin of an inhomogeneous Galactic halo (see e.g., discussion by  
\citealt{shen15,vandevoort15,kobayashi11}). 
Nevertheless, by adopting different ranges of stellar mass progenitors for the $r$ process, 
the GCE predictions show marginal variations for disk metallicities, and for the solar 
$s$ distribution.
\\
Even more puzzling is the origin of light neutron-capture elements in the 
early Galaxy, for which simple $r$-process residuals provide inadequate 
descriptions \citep{travaglio04}. 
Accordingly, a different treatment is required 

for Sr, Y and Zr. We derive an $r$ fraction of $\sim$10\% from observations of 
very metal-poor $r$-rich stars \citep{mashonkina14,roederer14}, under the hypothesis that 
these peculiar objects show the signature of a pure main $r$-process. 
An additional LEPP contribution was evoked by \citet{travaglio04} to explain the 
missing abundances of solar Sr, Y, and Zr. The flat [Sr,Y,Zr/Fe] trend observed
at low metallicities suggests that LEPP is a primary process, likely occurring in 
SNII with an extended range of mass progenitors compared to the main $r$ process.
As discussed by \citet{travaglio04}, we confirm that the solar LEPP-s described 
in Section~\ref{tests} and the metal-poor LEPP may originated in different 
stellar environments.

\begin{figure} 
\vspace{-30mm}
\includegraphics[angle=-90,width=26cm]{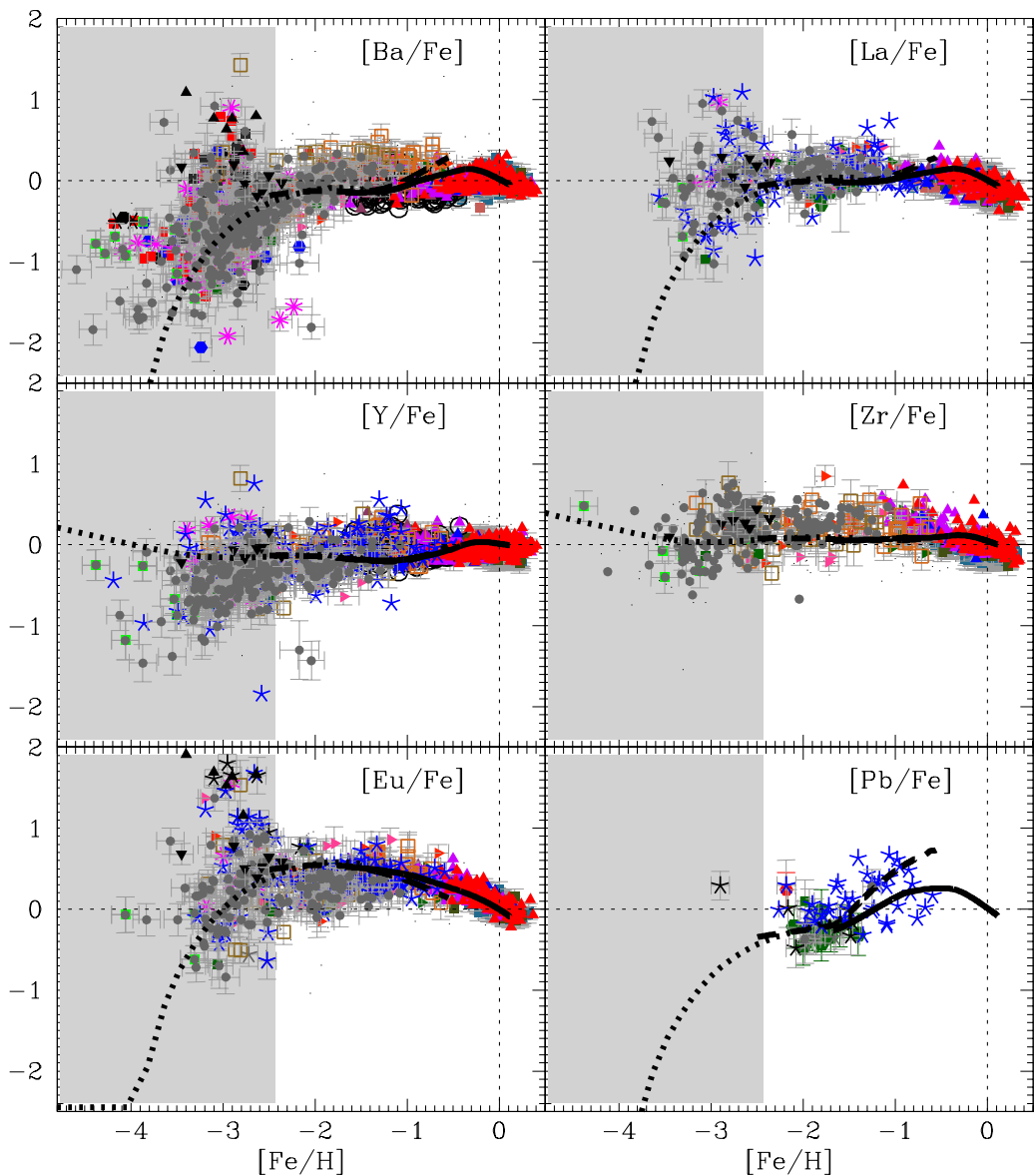}
\vspace{-25mm}
\caption{\label{Fig3} 
GCE s- and r-process contributions of [Ba/Fe], [La/Fe], [Y/Fe], [Zr/Fe], [Eu/Fe], 
and [Pb/Fe] versus [Fe/H]. 
Different lines correspond to GCE results for halo ($dotted$ $lines$), 
thick and thin discs ($dashed$ and $full$ $lines$, respectively)
obtained with a standard choice of the $^{13}$C-pocket (Table~\ref{Tab1}).
References and symbols for spectroscopic observations are given in the text.}
\end{figure}

\begin{figure} 
 \vspace{-40mm}
\includegraphics[angle=-90,width=26cm]{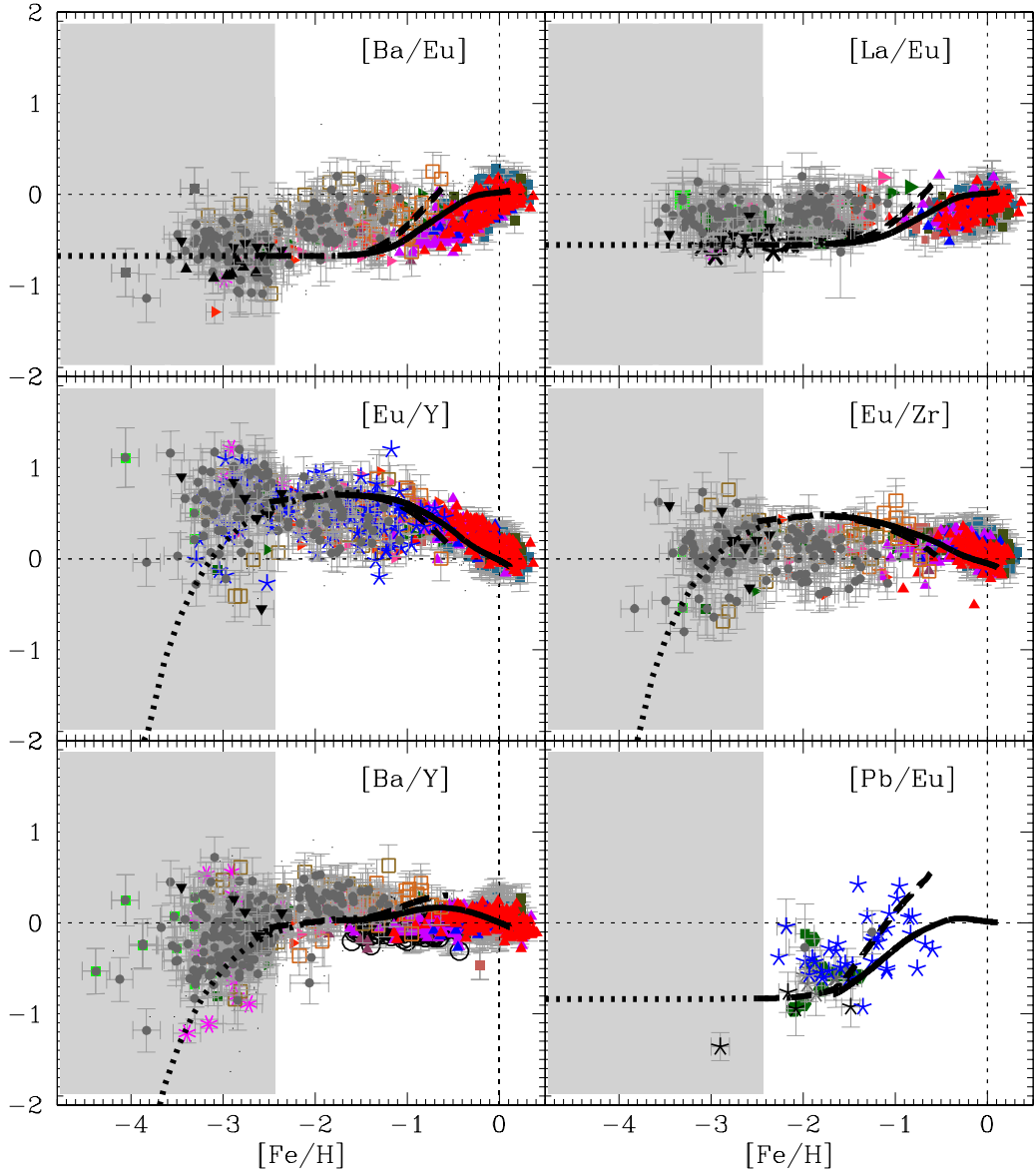}
\vspace{-25mm}
\caption{\label{Fig4} The same as Fig.~\ref{Fig3}, but for [Ba/Eu], [La/Eu], 
[Eu/Y], [Eu/Zr], [Ba/Y] and [Pb/Eu] versus [Fe/H]. }
\end{figure}

Figures~\ref{Fig3} and~\ref{Fig4} illustrate the 
Galactic evolution of [Ba/Fe], [La/Fe], [Y/Fe], [Zr/Fe], [Eu/Fe], 
[Pb/Fe] and their ratios [Ba/Eu], [La/Eu], [Eu/Y], [Eu/Zr], [Ba/Y],
[Pb/Eu] versus [Fe/H].
\\
Spectroscopic observations are taken from
Aoki \& Honda (2008; $dark$-$green$ $filled$-$diamonds$),
Nissen \& Schuster (2011; $big$-$empty$ $circles$),
Andrievsky et al. (2011; $red$ $squares$),
Aoki et al. (2013; $light$-$gray$ $squares$), 
Hansen et al. (2012; $brown$ and $sienna$ $empty$ $squares$ for dwarfs and giants, respectively), 
Hansen et al. (2014b; $black$ $asterisks$),
Roederer et al. (2010,2012,2014,2014a; $blue$ $stars$, $dark$-$green$ and $green$ $squares$, $gray$ $filled$-$circles$), 
Ishigaki et al. (2013; $green$, $orange$, and $violet$ $right$-$rotated$ $triangles$ for 
thick-disc, inner-halo and outer-halo stars, respectively), 
Mishenina et al. (2013; $light$-$blue$ and $sienna$ $squares$ for thin and thick disc stars, respectively), 
Cohen et al. (2013; $magenta$ $asterisks$), 
Yong et al. (2013; $blue$ $hexagons$), 
Bensby et al. (2014) and Battistini \& Bensby (2016; $red$, $blue$, $magenta$, $purple$ $triangles$ for 
thin disc stars, unclassified stars, thick-disc and halo stars, respectively), 
Mashonkina et al. (2014; $black$ $triangles$), 
Siqueira Mello et al. (2014; $black$ $down$-$rotated$ $triangles$).
Stars with [Ba/Fe] $\geq$ +0.6 and [Ba/Eu] $>$ 0 are excluded
(e.g., possible binaries as Ba stars, C-stars, CEMP-s stars).
The well studied sample of $r$-rich stars at [Fe/H] $<$ $-$2.2 (r-I and 
r-II stars, with [Eu/Fe] $\geq$ 1 and [Ba/Eu]$_{\rm r}$ $\sim$ $-$0.8;  
\citealt{mashonkina14,roederer14,siq14})
are peculiar objects likely born from a cloud polluted by a pure $r$-process
production event. 
They do not represent the average chemical evolution of our Galaxy. 
\\
GCE predictions account for the $s$-process, the $r$-process, and the 
LEPP contributions, in the halo, thick disk, and thin disk ($dotted$, 
$dashed$ and $full$ $lines$, respectively). AGB yields with a standard
$^{13}$C-pocket choice (first group of data in Table~\ref{Tab1}) are 
included.
The shaded area at [Fe/H] $\leq$ $-$2.5 indicates that homogeneous GCE models can not
properly represent the [El/Fe] scatter observed in the early Galaxy for neutron-capture
elements. 
It is noteworthy that low [Sr,Ba,Eu/Fe] abundances in metal-poor field stars  
may be affected by higher uncertainties owing to observational detection thresholds 
 \citep{roederer13,roederer14a}.

\begin{figure} 
\vspace{-10mm}
\includegraphics[angle=0,width=28cm]{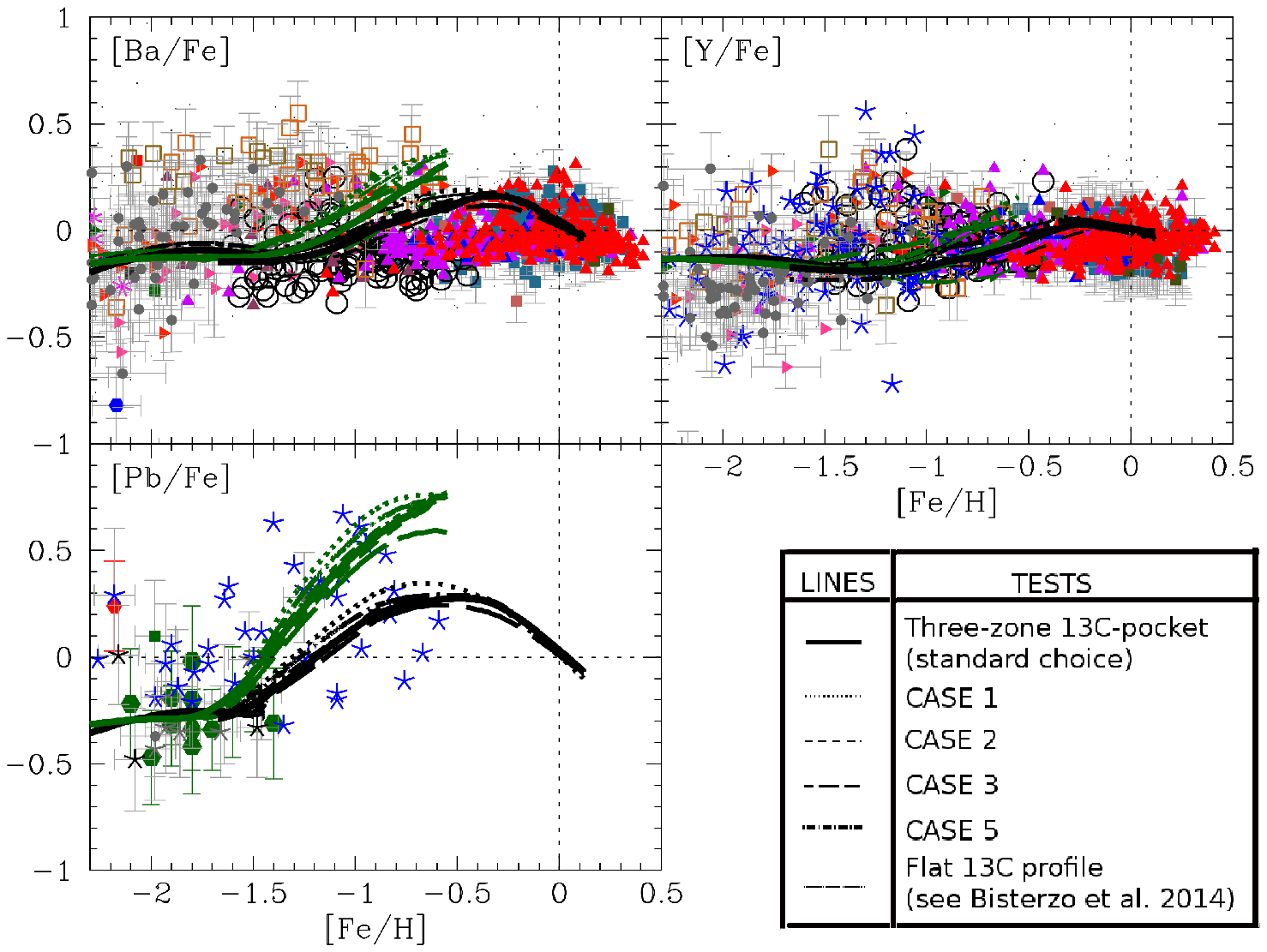}  
\vspace{-50mm}
\caption{\label{Fig5} GCE contributions of [Ba/Fe], [Y/Fe] and [Pb/Fe] 
versus [Fe/H]. Different lines correspond to GCE results for thick ($green$ $lines$)
and thin disc ($black$ $lines$) obtained with different $^{13}$C-pocket tests in 
Table~\ref{Tab1}. We also display the result of a flat $^{13}$C-pocket
profile obtained by excluding the two external $^{13}$C-rich zones of the pocket
(see \citealt{bisterzo14}).
Spectroscopic observations are the same shown in Fig.~\ref{Fig3}. } 
\end{figure}

In Figure~\ref{Fig5}, we illustrate the impact of AGB yields 
computed with different $^{13}$C-pocket tests selected from Section~2 
on GCE predictions of $s$-process
elements. We consider [Ba/Fe], [Y/Fe] and [Pb/Fe] versus [Fe/H]
as representative of the three $s$ peaks. Similar results are obtained 
for [La/Fe] and [Zr/Fe].
\\
As discussed in Section~\ref{tests}, [El/Fe] and their ratios are plausibly reproduced 
at [Fe/H] = 0. Variations up to $\Delta$[El/Fe]$\sim$$\pm$0.2 dex
are shown by GCE predictions of the three $s$ peaks in the metallicity range 
of $-$1.6 $<$ [Fe/H] $<$ 0.1. At [Fe/H] $<$ $-$1.6, the $s$ contribution is negligible, 
and marginal variations are expected.
Eu, which receives a dominant $r$-process contribution (94\%), is unaffected by  
different tests. 

The growing number of spectroscopic data currently suggests a possible 
underestimation of the observed [Ba/Fe] for [Fe/H] $\sim$ $-$2, 
independently of the $^{13}$C profile adopted in the pocket. 
For instance, the observed [Ba/Eu] ratio smoothly increases starting
from [Fe/H] $\sim$ $-$2.5 \citep{roederer14a,ishigaki13,hansen12}, 
about 1 dex earlier than predicted by our model ([Fe/H] $\sim$ $-$1.6). 
Despite a smaller dispersion is observed for [La/Fe], similar indications
are given by [La/Eu], which on average show an even flatter and higher trend
for [Fe/H] $<$ $-$2.
This trend appears a characteristic of the second $s$ peak  
rather than other $s$ elements (see, e.g., [Y,Zr/Fe]
and [Ba/Y] versus [Fe/H]). 
Otherwise, Galactic thin and thick disk dwarfs by  
Mishenina et al. (2013; little filled squares in the metallicity range of $-$1 $\leq$ 
[Fe/H] $\leq$ +0.3) show a delayed and steeper [Ba,La/Eu] increase, in agreement 
with our GCE $s$ process enrichment.
\\
Current AGB yields seem not give adequate hints to account for this behavior.
Within a GCE framework, several studies agree that the rise of the AGB $s$ process
occurs beyond [Fe/H] $\sim$ $-$1.6 \citep{matteucci09,kobayashi09}, given the fast
iron enrichment occurring in the early Galaxy.
However, an anticipated rise for the $s$-process was questioned by previous spectroscopic
studies \citep{simmerer04,hansen12}.

 Recent observational and theoretical studies have evidenced that an additional 
intermediate neutron-capture process (the $i$ process) may play an important 
role in the evolution of the Galaxy (see, e.g., peculiar chemical signatures
discussed by \citealt{roederer16,mishenina15a,dardelet15,lugaro15,jadhav13,liu14}).
Despite its existence was hypothesized 30 years ago by \citet{cowan77}, observational
hints have only recently been revealed.
The $i$ process occurs when protons are ingested in He-burning convective stellar 
regions to form $^{13}$C, which is the main source of neutrons via the ($\alpha$, n) 
reaction, driving neutron densities of $\sim$10$^{15}$ neutrons cm$^{-3}$.
The physical conditions leading to the $i$ process may be found in different stellar
environments, as super-AGB and post-AGB stars, He-core and He-shell flashes in 
low-metallicity low-mass stars, and massive stars 
\citep{cristallo09,herwig11,stancliffe11,jones16,wood15}.
Improved stellar models with the guidance from hydrodynamics simulations will help to
establish the impact of the $i$ process in the Milky Way.

\subsection{The impact of rotating-massive stars on GCE predictions of $s$-process
elements}\label{rot}

The important contribution of fast-rotating massive stars in the early Galaxy 
has been evidenced by several studies (see e.g., the review by \citealt{maeder12}). 
First investigations highlighted that rotational induced-mixing boost the production
of primary $^{14}$N in the H-burning shell, subsequently converted to $^{22}$Ne via 
2$\alpha$ captures in the He-burning core \citep{meynet06,hirschi07}. The large
amount of $^{22}$Ne primarily produced at low metallicities leads to an efficient 
nucleosynthesis of light $s$ elements, and may extend up to the barium peak 
\citep{pignatari08}. 
The recent studies by \citet{fris12} and \citet{fris16} show a large sensitivity 
of the first two $s$ peaks to both metallicity and rotation, with a [Sr/Ba] predicted
ratio that covers more 2 dex in most metal-free models. Rotating low-metallicity 
massive stars are a promising site to explain the [Sr/Ba] dispersion observed in the 
halo. \citet{chiappini11} first advanced the hypothesis that metal-poor 
fast-rotating massive stars may offer a plausible explanation for the large [Ba/Y] 
dispersion observed in the early Universe. Later on, the inhomogeneous chemical 
evolution model by \citet{cescutti13} provided a quantitative estimation in support
of this hypothesis.

The aim of this Section is to test the effect of weak $s$ process yields from 
fast-rotating massive stars on GCE predictions of $s$-process
elements. Specifically, new yields by
\citet{fris16} are analyzed in the framework of our GCE model in order to explore
the impact on the solar and metal-poor LEPP mechanisms.
\\
\citet{fris16} provide a grid of 
four stellar masses ($M$ = 15, 20, 25 and 40 $M_\odot$) and three metallicities 
($Z$ = $Z_\odot$, 10$^{-3}$, 10$^{-5}$), computed with a full s-process network. 
We have linearly interpolated the weak $s$ yields between the mass
range of 15 $<$ $M$/$M_\odot$ $\leq$ 40 and the metallicity range.
The authors compare non-rotating and standard-rotating models ($v_{ini}$/$v_{crit}$ 
= 0.4, where the critical velocity $v_{crit}$ assumed by the authors corresponds to an
average equatorial velocity of $\sim$200 km s$^{-1}$ on the main sequence at 
$Z_\odot$, and increases to $\sim$400 km s$^{-1}$ at low metallicity, see \citealt{fris12}).
They also provide additional yields on the 25 $M_\odot$ model down to $Z$ = 
10$^{-7}$, with fast-rotation rates ($v_{ini}$/$v_{crit}$ = 0.5 and 0.6, with 
$v_{crit}$ up to $\sim$590 km s$^{-1}$).
For fast-rotating low-metallicity models of 25 $M_\odot$, the authors discussed a further
test in which the $^{17}$O($\alpha$, $\gamma$) rate is divided by 
a factor of ten, in better agreement (within the uncertainties) with the recent
measurement by \citet{best13}.
\\
Although only pre-explosive SN yields are given by \citet{fris16}, they are suited for 
GCE studies because strong variations on the total yields of s-process nuclei are not 
expected after SN explosion (see e.g., \citealt{tur09}).

We have considered three tests: \\
-- a first test with non-rotating weak $s$-process yields (label $NR$; corresponding to models 
`$s0$' in Table~1 by \citealt{fris16}), which provides a more exhaustive
analysis of the `classical' weak-$s$ process contribution included in the GCE predictions 
presented in Section~\ref{z};\\
-- a second test with standard-rotating weak $s$-process yields (label $SR$; corresponding to models 
`$s4$' in Table~1 by \citealt{fris16}, with $v_{ini}$/$v_{crit}$ = 0.4); \\
-- the same as the second test, but with fast-rotating yields ($v_{ini}$/$v_{crit}$ = 0.6; $FR$) 
coupled with the $^{17}$O($\alpha$, $\gamma$) rate divided by a factor of ten for the 25 $M_\odot$ 
model at $Z$ = 10$^{-5}$ and 10$^{-7}$ (corresponding to models `$s6b$' in Table~1 by \citealt{fris16}). 
We applied the same scaling factors to 40 $M_\odot$ models.

\begin{figure} 
\includegraphics[angle=-90,width=15cm]{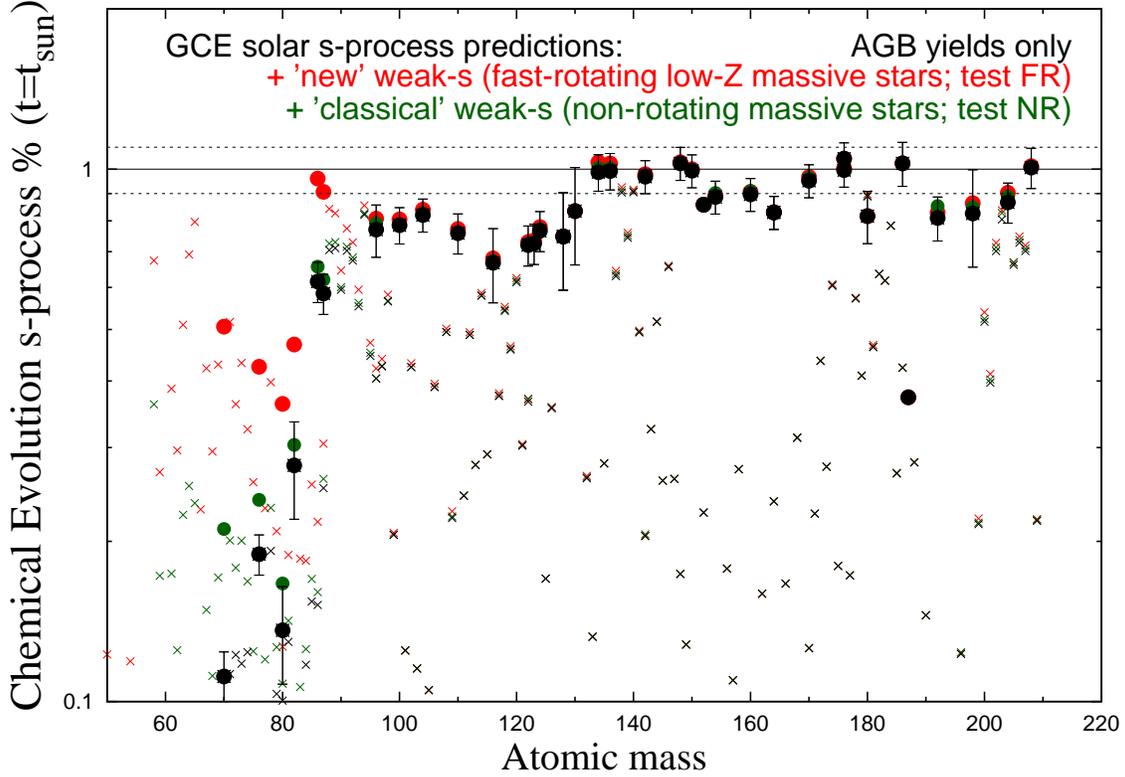}  
\caption{\label{Fig6} Effect of rotating-massive stars on GCE predictions of $s$-process
elements. Black symbols stand for the standard GCE results obtained 
in Fig.~\ref{Fig1}. GCE predictions computed with non-rotating weak $s$ yields
are represented with $green$ $symbols$ (test $NR$). GCE results that include 
fast-rotating models and the $^{17}$O($\alpha$, $\gamma$) rate divided by a 
factor of ten are displayed with $red$ $symbols$ (test $FR$). 
See text for more details.} 
\end{figure}

In Fig.~\ref{Fig6}, the impact of new weak $s$ yields is shown for the solar abundance 
distribution of $s$ isotopes.
Non-rotating weak $s$ yields mainly increase the solar abundances of isotopes with $A$ $\la$ 80, 
with a small contribution to Sr, Y and Zr. This is in agreement with previous `classical' analyses, 
which have been obtained starting from the solar weak $s$ contribution and following a 
secondary-like behavior by decreasing the metallicity.
Fast-rotating models produce an evident increase of isotopes lighter than $A$ $\la$ 90, 
with an additional contribution of $\sim$30\% to $^{86,87}$Sr, and $\sim$10\% to $^{89}$Y, $\sim$5\%
to $^{90,91,92}$Zr among non $s$-only isotopes. This corresponds to a weak $s$ contribution 
from fast-rotating massive stars of about 17\%, 10\% and 5\% to solar Sr, Y, Zr, respectively.
In the atomic mass region between 90 $<$ $A$ $<$ 140, $s$-only isotopes are marginally 
affected, with variations within the solar uncertainties. This suggests that fast-rotating 
massive stars may only partially account for the solar LEPP, being inefficient for $A$ $>$ 90.

\begin{figure} 
\vspace{-15mm}
\includegraphics[angle=-90,width=26cm]{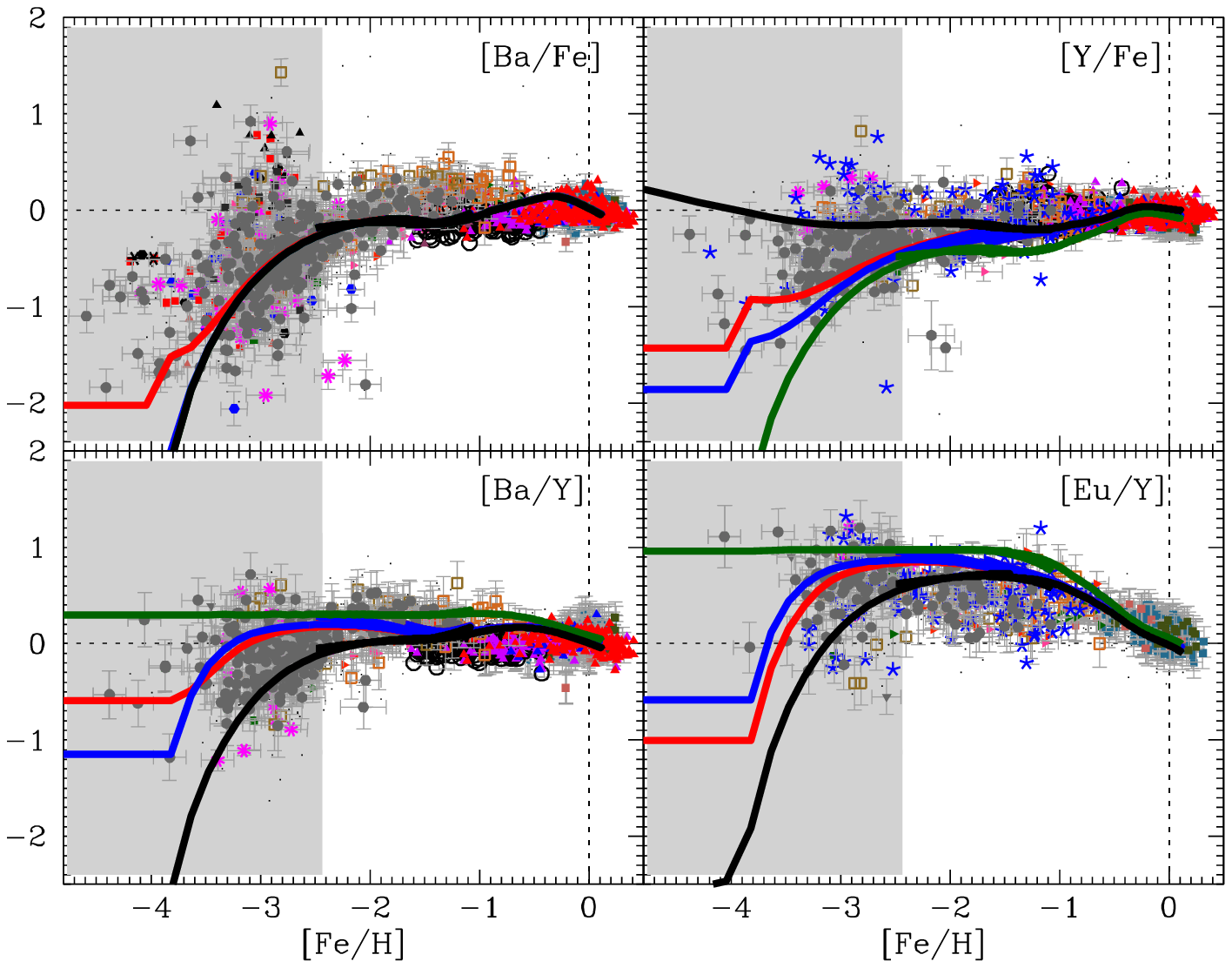}  
\vspace{-70mm}
\caption{\label{Fig7} Observations [Ba/Fe], [Y/Fe] and the ratios [Ba/Y], [Eu/Y]
versus [Fe/H] are compared with GCE predictions computed with different weak $s$
process contributions.
Black lines represent the same GCE predictions shown in Fig.s~\ref{Fig3} and~\ref{Fig4}, 
in which the `classical' weak-$s$ process is included.
Other lines show GCE results obtained by testing the weak-$s$ process 
yields recently provided by \citet{fris16} for massive stars with different
initial rotation rates: non-rotating ($green$ $lines$; test $NR$) and standard-rotating models 
($blue$ $lines$; test $SR$), and fast-rotating models coupled with the $^{17}$O($\alpha$, 
$\gamma$)$^{21}$Ne rate divided by a factor of ten ($red$ $lines$; test $FR$). }
\end{figure}

By considering the influence of the fast-rotating weak-s process over the whole Galactic
enrichment, we display in Fig.~\ref{Fig7} the impact on [Ba/Fe], [Y/Fe], [Ba/Y] and [Eu/Y] 
versus [Fe/H], selected as most representative.
Black lines account for both the `classical' weak-$s$ process and 
for the estimated LEPP. 
Other lines are computed by setting the LEPP contribution to zero, and by including
the three tests listed above for the weak-$s$ process. 
We adopt our standard treatment described in Section~\ref{z} for the $r$ contribution.
The $r$ process dominates the [Ba/Fe] predictions for $-$3 $\leq$ [Fe/H] $\la$ $-$1.5.
The impact of fast-rotating models only affects very metal-poor GCE predictions (see
$red$ $line$ at [Fe/H] $<$ $-$3).
As expected, Eu (which is not shown in Figure) does not present variations.
For [Y/Fe], different initial rotation rates (represented by $green$, $blue$, and 
$red$ lines in Fig.~\ref{Fig7}) seem to cover a more extended range of stellar 
observations at [Fe/H] $<$ $-$3. This also partly explains the observed [Eu/Y] and 
[Ba/Y] dispersions. 
It is noteworthy that fast-rotating massive stars offer a 
promising nucleosynthesis site for the metal-poor LEPP. For instance, GCE predictions 
that account for fast-rotating weak-s process yields ($red$ line) move towards the original
model with the estimated LEPP ($black$ line). 
Although our homogeneous GCE model can not provide a realistic representation of 
the observed s-element dispersion, the averaged GCE predictions shown in Fig.~\ref{Fig7}
are in agreement with previous results presented by \citet{cescutti13}, who indicated
that low-metallicity fast-rotating massive stars may reproduce the [Sr/Ba] inhomogeneities
observed in extremely metal-poor stars.
\\
Note that large uncertainties affect massive-star models, as discussed
by \citet{fris16}: on the nuclear point of view, owing to the uncertain $^{17}$O+$\alpha$ 
and $^{22}$Ne+$\alpha$ rates, and on the modeling point of view, mainly because of the 
treatment of the physics associated to magnetic fields and rotation-induced mixing. 
In the framework of GCE models, a more detailed grid of stellar yields may help to better 
assess an upper limit for the fast-rotating weak-s process contribution.
\\
As evidenced by different GCE approaches, the chemical enrichment of the stellar halo 
 is likely the result of a merger of different (primary) processes, including 
fast-rotating weak-s process and $r$ processes (e.g., 
\citealt{ishimaru99,argast04,cescutti08,matteucci14};
see also Section~\ref{z}).

\section{Conclusions}\label{conclusions}

We have implemented the study on GCE predictions of $s$-process
elements presented by \citet{bisterzo14} by
including recent pivotal suggestions provided in literature. 

The main result is that the predicted chemical compositions in the interstellar medium at any epoch 
are grossly independent of the particular $^{13}$C profile adopted inside the pocket,
once a different weighted average of $^{13}$C-pocket strengths is considered to 
reproduce solar $^{150}$Sm (within 5\% accuracy).
\\
The impact of AGB uncertainties on GCE computations may be partially reduced by assuming
a range of $^{13}$C-pocket strengths, as indicated by observational constraints. 
Actually, this approach allows us to obtain more accurate information about the complementary 
contribution from other nucleosynthesis processes that may compete over the Galactic history.
In the framework of these $^{13}$C-pocket prescriptions, we confirm that the 
additional LEPP-s mechanism invoked by \citet{travaglio04} is 
required to represent the observed solar distribution within the estimated solar uncertainties. 
 In this view, only 3D or hydrodynamical AGB modeling may shed light on the $^{13}$C-pocket 
scenario.
\\
Updated GCE calculations accounting for recent fast-rotating metal-poor weak $s$-process
yields by \citet{fris16} may partially solve the missing contribution to the solar Sr-Y-Zr
abundances. 
The [Sr/Ba] dispersion observed in the early Galaxy may be reconciled with the contribution 
of metal-poor massive stars with different rotational velocity, in agreement with previous 
results by \citet{chiappini11} and \citet{cescutti13}.
A more detailed grid of stellar yields would help to assess the total $s$ 
contribution. From the present analysis, a combination of different $r$ and $s$ process 
components is the most promising scenario to explain the solar and the metal-poor LEPP.
\\
In this view, additional processes have to be invoked to explain the observed trends.
The intermediate neutron-capture process ($i$ process) may play a decisive 
role in the nucleosynthesis of the Galaxy (see \citealt{roederer16,mishenina15a}, and 
references therein).

This issue deserves a more comprehensive analysis in order to establish whether the 
present discrepancies between GCE predictions and the analyzed spectroscopic sample
are confirmed by other elements.
Future investigations with a more extended and homogeneous data-set of neutron-capture 
elements (e.g., Cu, Ga, Ge, among weak $s$ process elements; Mo, Ru, Ag, Pd, for weak
$r$ process elements; Ce, which is expected to behave as Ba) are planned to distinguish 
between different nucleosynthesis processes.

The improved quality and statistics of the ongoing surveys (e.g., GAIA-ESO, 
SEGUE, RAVE, APOGEE, HERMES-GALAH) will also assure a more solid picture of the Galactic
sub-structures. 
First studies by \citet{edvardsson93}, \citet{reddy03} and \citet{bensby03}
evidenced that stars belonging to the thin and thick discs have different ages, kinematics, 
locations on the Galactic plane, and chemical properties.
Although a complex image of the Galactic structure emerges from the most recent 
investigations (e.g., perturbation, accretion, merging or heating processes, radial
migrations, gas flows; see the review by \citealt{rix13}), the main features observed
in the solar neighborhood are confirmed by several surveys. For instance, it has been 
well established in literature that various $\alpha$-elements show clear separate behaviors
in the thick and the thin disc. 
\\
In this view, a detailed analysis of our GCE prescriptions is required in order to 
provide an improved representation of the chemical abundances observed in thick/thin disc stars
(see, e.g., [O/Fe] vs [Fe/H] in Fig.~1 by \citealt{bisterzo14}).
\\
A preliminary study of the main GCE ingredients (e.g., star formation rate, initial 
mass function, delay-time distribution function for SNIa, and the present uncertainties 
affecting SNIa stellar yields) highlights that the theoretical interpretation of the 
thick/thin disc observed trends is largely affected by the poorly known SNIa scenario (see,
e.g., \citealt{maoz14,ruiz14,travaglio15,marquardt15,hillebrandt13,ruiter09}, and 
references therein).
Specifically, our GCE thick/thin disc predictions may be improved by assuming a factor 
of 2 uncertainty in the SNIa stellar yields adopted so far \citep{travaglio05}, 
coupled with an updated treatment of the delayed-time distribution (DTD) function (as 
suggested by \citealt{kobayashi98,greggio05,matteucci09,kobayashi15}), in which we 
assume a dominant SNIa contribution starting from [Fe/H] $>$ $-$1.
\\
Additionally, distinct enrichment histories of the thick/thin disc components may be obtained
by adopting a different star formation rate, with a delayed thin disc phase 
(as hypothesized by the two-infall model by \citealt{chiappini97}). To this purpose,
we will provide a comprehensive investigation of several GCE ingredients against a large 
number of elements in a forthcoming paper. The underway study of SNIa yields for an extended 
metallicity grid will further constraint the disc chemical enrichment.
New SNIa inputs,
combined with high-resolution studies of several chemical elements and the reliable age 
determination expected from astroseismology (CoRoT and Kepler, \citealt{anders16,stello15}),
will provide a real breakthrough in the understanding of the Galactic history.

It is noteworthy that 
we will extend the analysis to young open clusters, which are mandatory to understand 
the chemical evolution of neutron-capture elements in the Galactic disc 
\citep{maiorca12,mishenina15a}. 

We plan to investigate in detail these topics in a forthcoming paper.

\acknowledgments

 We acknowledge the anonymous Referee for offering helpful suggestions.
We thank D. Yong for precious information about spectroscopic observations.
The present work has been supported by JINA (ND Fund \#202476). Numerical calculations 
have been sustained by B2FH Association (http://www.b2fh.org/).

\end{document}